\documentclass[12pt,preprint]{aastex}

\newcommand{\myemail}{ asai@astro.s.chiba-u.ac.jp}

\shorttitle{MHD Simulations of Cold Fronts}
\shortauthors{Asai, Fukuda, \& Matsumoto}

\begin{document}

\title{Magnetohydrodynamic Simulations of the Formation of Cold Fronts
	in Clusters of Galaxies including Heat Conduction}

\author{N. Asai\altaffilmark{1}, N. Fukuda\altaffilmark{2}, 
and R. Matsumoto\altaffilmark{3}}

\altaffiltext{1}{Graduate School of Science and Technology, 
Chiba University, 1-33 Yayoi-cho, Inage-ku, Chiba 263-8522, Japan;
\myemail}

\altaffiltext{2}{Department of Computer Simulation,
Faculty of Informatics, Okayama University of Science,
1-1 Ridai-cho, Okayama 700-0005, Japan; fukudany@sp.ous.ac.jp}

\altaffiltext{3}{Department of Physics, Faculty of Science, 
Chiba University, 1-33 Yayoi-cho, Inage-ku, Chiba 263-8522, Japan; 
matumoto@astro.s.chiba-u.ac.jp }

\begin{abstract}
Recent {\it Chandra} observations of clusters of galaxies 
revealed the existence of a sharp ridge in the X-ray surface brightness 
where the temperature drops across the front. 
This front is called the cold front. We present 
the results of two-dimensional magnetohydrodynamic simulations
of the time evolution of a dense subcluster plasma moving in a cluster of 
galaxies. 
Anisotropic heat conduction along the magnetic field lines is included.
In the models without magnetic fields, the numerical results 
indicate that the heat conduction from the hot ambient plasma 
heats the cold dense plasma of the subcluster and diffuses 
out the cold front. When magnetic fields exist in a cluster of galaxies, 
however, cold fronts can be maintained because the heat conduction across the
magnetic field lines is suppressed. 
We found that, even when the magnetic fields in a cluster of galaxies are 
disordered, heat conduction across the front is restricted 
because the magnetic field lines are stretched along the front. 
Numerical results reproduced the X-ray intensity distribution 
observed in the A3667 cluster of galaxies.

\end{abstract}

\keywords{conduction---MHD---galaxies: cluster: magnetic fields---
X-ray: galaxies---plasmas}

\section{INTRODUCTION}
High-resolution {\it Chandra} observations have revealed various 
substructures in the hot X-ray emitting plasmas in clusters of galaxies. 
For example, sharp discontinuities of X-ray surface brightness 
distributions have been found around subclusters moving in a cluster 
of galaxies such as A2142 \citep{mar00} and A3667 
(Vikhlinin, Markevitch, \& Murray 2001b).
The temperature drops sharply across this boundary between the 
gravitationally confined, cold dense plasmas in the moving subcluster 
and the hot, less-dense ambient plasma.
This sharp discontinuity is called the cold front. 
It is not a shock front, because {\it Chandra} images of the 
X-ray surface brightness show that the temperature decreases 
in the denser side.

\citet{ett00} suggested that the sharp temperature gradient 
observed by {\it Chandra} in A2142 requires the heat 
conductivity near the front to be reduced by several 
orders of magnitude from the classical Spitzer value \citep{spi62}, 
$\kappa_{\rm Sp}= 5 \times 10^{-7} T^{5/2}$ 
$\, {\rm erg \, s^{-1} \, cm^{-1} \, K^{-1}}$.

Based on the observations of A3667, 
Vikhlinin, Markevitch, \& Murray (2001a) suggested that 
turbulent, intergalactic magnetic fields stretched along 
the cold front reduce the growth rate of the Kelvin-Helmholtz (K-H) 
instability, which possibly destructs the front. 
Furthermore, heat conduction across the magnetic fields 
can be suppressed. It has been recognized that heat conduction 
in a cluster of galaxies can be very efficient (e.g., \citealt{tak77}) 
because the Coulomb mean free path, 
$l_{\rm c} \sim 4 \, (kT / 5 \, {\rm keV})^{2}(n / 10^{-3}{\rm cm^{-3}})^{-1} 
\, {\rm kpc}$, is large. Here, $n$ is the number density. 
The time required for heat to diffuse by conduction across a length  
$L$ is given by $\tau_{\rm Sp} \sim \rho L^{2}/ \kappa_{\rm Sp}
\sim 10^{7} (kT/ 5 \, {\rm keV})^{-5/2}
(n / 10^{-3} \, {\rm cm^{-3}})(L/100 \, {\rm kpc})^{2} \, {\rm yr}$, 
where $\rho$ is the density.
However, if magnetic fields exist, the characteristic length of the 
heat exchange across magnetic field lines is reduced to the Larmor radius, 
$r_{\rm L} = 2500 \, (B/1 \, {\rm \mu G})^{-1}$ $(T/5 \, {\rm keV})^{1/2} 
\, {\rm km}$.
Since the typical magnetic field strength in a cluster of galaxies is 
$\sim {\rm \mu G}$ (e.g., \citealt{kro94,car02}), 
the heat conductivity across magnetic field lines is drastically reduced.

In this Letter, we present the results of numerical simulations of the 
formation of cold fronts. These simulations include the effects of 
heat conduction.
We adapted the simulation code, originally developed for
solar flare simulations \citep{yok97,yok01}, 
which includes anisotropic heat conduction along magnetic field lines.

\section{SIMULATION MODELS}

We simulated the time evolution of the cluster plasma in 
a frame comoving with the subcluster. The basic equations are as follows, 

\begin{equation}
\frac{\partial \rho}{\partial t} + \mbox{\boldmath $\nabla$ $\cdot$}
 (\rho \mbox{\boldmath $v$}) = 0,
\end{equation}

\begin{equation}
\rho \left[
\frac{\partial \mbox{\boldmath $v$}}{\partial t} 
+ (\mbox{\boldmath $v$ $\cdot$ $\nabla$}) 
\mbox{\boldmath $v$}
\right] =
-\mbox{\boldmath $\nabla$}{\it p} + 
\frac{(\mbox{\boldmath $\nabla$} \times 
\mbox{\boldmath $B$}) \times \mbox{\boldmath $B$}}{4 \pi} 
- \rho \mbox{\boldmath $\nabla$} \psi, 
\end{equation}

\begin{equation}
\frac{\partial \mbox{\boldmath $B$}}{\partial t} 
= \mbox{\boldmath$\nabla$} \times (\mbox{\boldmath $v$} 
\times \mbox{\boldmath $B$}),
\end{equation}

\begin{equation}
\frac{\partial}{\partial t} 
\left[
\frac{1}{2} \rho v^{2} 
+ \frac{B^{2}}{8 \pi} + \frac{p}{\gamma -1}
\right]
+ \mbox{\boldmath$\nabla$$\cdot$}
\left[
\left(\frac{1}{2} \rho v^{2} + \frac{\gamma p}{\gamma -1}
\right) \mbox{\boldmath$v$} + \frac{1}{4 \pi} (-\mbox{\boldmath $v$} 
\times \mbox{\boldmath$B$}) 
\times \mbox{\boldmath $B$} - \kappa_{\parallel}  
\mbox{\boldmath $\nabla_{\parallel}$} T
\right]
= - \rho \mbox{\boldmath $v$ $\cdot$ $\nabla$} \psi,
\end{equation}
where $\rho$, $\mbox{\boldmath $v$}$, $p$, $\mbox{\boldmath $B$}$, and 
$\psi$ are the density, velocity, pressure, magnetic fields, 
and gravitational potential, respectively. We use the specific heat ratio 
$\gamma = 5/3$. The subscript $\parallel$ denotes the components parallel 
to the magnetic field lines. We assume that heat is conducted only 
along the field lines.
We solved the magnetohydrodynamic (MHD) equations by a modified 
Lax-Wendroff method with artificial viscosity. 
The heat conduction term in the energy equation is 
solved by the SOR method (see \citealp{yok01} for details).
The units of length, velocity, density, pressure, temperature, and time 
in our simulations are 
$r_{0}=250 \, {\rm kpc}$, 
$v_{0}=1000 \, {\rm km \, s^{-1}}$,
$\rho_{0}=5 \times 10^{-27} \, {\rm g \, cm^{-3}}$, 
$p_{0} = 3 \times 10^{-11} \, {\rm erg \, cm^{-3}}$, 
$T_{0}= 4 \, {\rm keV}$, and 
$t_{{0}}=r_{0}/v_{0}= 3 \times 10^{8} \, {\rm yr}$,
respectively. We adopted two-dimensional Cartesian coordinates $(x, y)$
by assuming translational symmetry in the $z$-direction.
The typical size of the simulation box and the number of grid points are 
$2.5 \, {\rm Mpc} \times 2.5 \, {\rm Mpc}$, and 
$(N_{x}, N_{y}) = (1201,1201)$, respectively.

Figure \ref{fig1} shows the initial density distribution. 
We assume that the subcluster is a sphere having a radius 
$r=500 \, {\rm kpc}$, 
and it is isothermal with temperature $kT_{\rm in} = 4 \, {\rm keV}$.
We assume that the gravitational potential at $r = (x^{2}+y^{2})^{1/2}$ 
is given by $\psi (r) = (c_{\rm s}^{2}/\gamma) \ln[1+(r/r_{\rm c})^{2}]$,
where we use the core radius $r_{\rm c} = 290 \, {\rm kpc}$, 
and the sound speed inside 
the subcluster $c_{\rm s \, in} = 1000 \, {\rm km \, s^{-1}}$, 
where the subscript $in$ denotes the values inside the subcluster.
The density distribution is determined by the hydrostatic equation 
$\mbox{\boldmath$\nabla$}p = - \rho \mbox{\boldmath $\nabla$} \psi$, as 
$\rho_{\rm in} (r) = \rho_{\rm c} / [1 + (r/ r_{\rm c})^{2}]$,
where we adopt 
$\rho_{\rm c}= 10^{-26} \, {\rm g \, cm^{-3}}$. 
We also assume that the subcluster is surrounded by 
hot $(kT_{\rm out} = 8 \, {\rm keV})$, less-dense 
$(\rho_{\rm out} = 0.25 \, \rho_{0})$ 
plasma, where the subscript $out$ denotes the values outside the subcluster. 
The gravitational potential is taken to be fixed throughout
the simulation.

In order to simulate the moving subcluster in a cluster of galaxies, 
we assume that ambient plasmas initially have a uniform speed with 
Mach number $M = v_{x}/c_{\rm s \, out} = 1$, 
where $c_{\rm s \, out}$ is the sound speed of the ambient plasmas. 
The Mach number 
with respect to the sound speed inside the subcluster is 
$ M^{\prime}= v_{x}/c_{\rm s \, in}= \sqrt{2}$. 

Table \ref{tbl-1} shows the model parameters.
Models H and HC are without magnetic fields. In models M and MC,
we assume uniform magnetic fields parallel to the $y$-axis. In these models, 
the initial ratio of the ambient gas pressure to the magnetic pressure 
($\beta= p_{\rm gas} / p_{\rm mag}$) is taken to be $\beta_{0} = 100$, 
which corresponds to the magnetic field strength 
$ B \sim 3 \, {\rm \mu G}$.

In models HC, MC, and MTC we included heat conduction. 
When the magnetic field does not exist, heat is conducted isotropically.
When the Coulomb mean free path $l_{\rm c}$ becomes comparable to 
or greater than the temperature scale height 
$l_{T} = T /|\mbox{\boldmath $\nabla$} T|$, the heat flux $Q$ 
is saturated as 
$Q_{\rm sat}= - [l_{T}/(4.2 \, l_{\rm c})] \kappa_{\rm Sp} 
\mbox{\boldmath $\nabla$} T$ when $l_{T} < 4.2 \, l_{\rm c}$ 
(e.g., \citealt{cow77}). 
In our initial model, since $l_{\rm c} \simeq 10 \, {\rm kpc}$ and 
$l_{T} < 6 \, {\rm kpc}$ around the cold front,
$Q_{\rm sat}$ at the front is smaller than 15\% of the classical value.
However, this reduced heat flux is large enough to smear the cold front 
within $10^{7} \, {\rm yr}$ because the conduction time scale across the 
cold front is 
$\tau_{\kappa} \sim (4.2 \, l_{\rm c}/ l_{T}) \, \rho \, {l_{T}}^{2} 
/ \kappa_{\rm Sp} = 2 \times 10^{5} \, (kT/ 5 \, {\rm keV})^{-5/2} 
(n / 10^{-3} \, {\rm cm^{-3}}) (l_{T}/6 \, {\rm kpc})
(4.2 \, l_{\rm c}/42 \, {\rm kpc}) \, {\rm yr}$.

When the magnetic field exists, heat is conducted only in the 
direction parallel to the magnetic field lines. 
The conductivity parallel to the field lines is  
$\kappa_{\parallel} = 5 \times 10^{-7} \, T^{5/2}$ $ \, {\rm erg \, s^{-1} 
\, cm^{-1} \, K^{-1}}$. 
Conversely, the conductivity across the field lines is $ \kappa_{\perp} = 0$. 
The conduction time scale along the field lines is 
$\tau_{\kappa} \sim \rho L^{2}/ \kappa_{\parallel}
\sim 10^{7} \, (kT/ 5 \, {\rm keV})^{-5/2} (n / 10^{-3} \, {\rm cm^{-3}})$
$(L/100 \, {\rm kpc})^{2} \, {\rm yr}$. 

In model MTC, we simulated a subcluster moving in an ambient plasma with initially disordered magnetic 
fields, and we included the effects of heat conduction.
In this model, we superposed magnetic fields created by thin currents 
flowing in the $z$-direction. The location and strength of the thin currents are 
chosen randomly. In this particular model, the size of the simulation box is 
$7.5 \, {\rm Mpc} \times 2.5 \, {\rm Mpc}$, and the number of grid 
points is $(N_{x}, N_{y}) = (901, 301)$. 

The left boundary at $x = -5$ is taken to be a fixed boundary, except in 
model MTC. 
In model MTC, the magnetic fields at the left boundary are extrapolated from 
magnetic fields 
in internal points. Other quantities are fixed at the left boundary. 
The effects of the left boundary condition is negligible in the time 
range we report in this Letter.  
Other boundaries are free boundaries where waves can be transmitted.

\section{RESULTS}

\subsection{Effects of Magnetic Fields on Heat Conduction}\label{sec:mc}

Figure \ref{fig2} shows the distributions of density, temperature and 
X-ray intensity for model MC (upper panels) and model HC (lower panels) 
at $t=3.0$ (0.9 Gyr). 
Model MC includes the effects of anisotropic heat conduction; that is, 
heat is conducted only in the direction parallel to the magnetic field lines. 
The X-ray intensity distribution is visualized from 
the simulation results as the logarithm of $\rho^{2}$.
In model HC with isotropic heat conduction, 
the subcluster plasma is heated by the ambient hot plasma 
and evaporates rapidly, 
because heat is conducted from the ambient plasma to the cold subcluster 
plasma. Thus, the sharp front of the temperature distribution disappears.
We also carried out simulations that included the saturation effects of heat 
flux, and we confirmed that the effects are negligible. 
In model MC, 
the heat conduction and plasma diffusion across the front are 
suppressed because the magnetic field lines are parallel to the cold front.
The X-ray intensity distribution in model MC shows 
a steeper gradient at the leading edge of the subcluster than that in 
model HC. 
The numerically obtained X-ray intensity distribution of model MC 
reproduces the overall structures of X-ray intensity observed in A3667.
At $t=3.0$, the magnetic field strength near the front increases 
by a factor of 8, and $\beta$ decreases by a factor of 100 from the 
initial state.
In this model, the lifetime of the cold front is 
more than $10^{9} \,{\rm yr}$. 

Numerical results also reproduced the bow shock upstream of the cold front.
The interaction between the inflowing hot ambient plasma and the 
cold subcluster plasma drives another shock propagating inside the subcluster. 
This shock is located at $(x,y)=(2.5,0)$ in the upper panels of Figure \ref{fig2}. 
The shock propagating inside the subcluster initially smears the cold front by 
shock heating. Subsequently, adiabatic expansion 
following the shock compression cools the subcluster plasma. 
This cold plasma is pulled back by gravity and the cold front is recovered.

\subsection{Comparison with Models without Heat Conduction}\label{sec:m}

Figure \ref{fig3} shows the temperature distribution at $t=3.0$ in 
models M (left) and H (right). 
In these models, heat conduction is not included. 
By comparing model H with HC, we confirm that 
when the magnetic field does not exist, cold fronts 
can be maintained only if the heat conductivity is small enough. 

The results of model H indicate that the K-H instability 
grows at the contact discontinuity between the cold, dense subcluster 
plasma and the hot, less-dense ambient plasma. The growth of 
the K-H instability is not prominent near the stagnation 
point $(x,y) \sim (-1,0)$, but the K-H vortices grow 
around $(x,y) \sim (2,\pm 2)$. 
The K-H instability does not disrupt the cold front, because the velocity 
difference inside and outside of the contact surface is small in the 
upstream side of the dense subcluster.
In model M, the magnetic tension force along the front 
suppresses the K-H instability, even in the sides of the dense 
subcluster, 
where the velocity shear is large. 
Except for this stability for the K-H instability, 
the numerical results of model M are similar to these of model H.

\subsection{Effects of Disordered Magnetic Fields}\label{sec:mtc}
 
Figure \ref{fig4} shows the results of model MTC, 
in which we initially assumed disordered magnetic fields. 
The top and bottom panels show the time evolution of the density and 
temperature distributions at $t=0$ and $t= 3.0$ ($0.9 \, {\rm Gyr}$), 
respectively. 
The white curves and arrows show the magnetic field lines and velocity vectors, 
respectively. 
Numerical results show that the tangled magnetic field lines in the vicinity 
of the front are compressed and stretched along the front. 
The temperature distributions 
show that heat is conducted along the tangled field lines.
As a result, the low-temperature subcluster plasma is heated and diffuses 
along the field lines. 
Since the tangled magnetic field lines are stretched along the front, 
suppression of heat conduction across the field lines enables 
the existence of cold fronts.

\section{DISCUSSION}

\citet{hei03} carried out hydrodynamical simulations of 
the interaction of cold subcluster plasma and hot ambient matter. 
They suggested that a cold front can be formed after the passage of 
shock waves through the subcluster plasma, when the cold plasma, initially 
pushed forward by the shock, falls back. However, in their simulations 
heat conduction and magnetic fields were neglected.
In this Letter, we presented the results of numerical simulations 
which included both magnetic fields and heat conduction. 
We confirmed that, since the Spitzer conductivity is very large 
in hot ambient plasma, cold fronts cannot be maintained longer 
than $0.3 \, {\rm Gyr}$, unless the heat conduction across 
the front is suppressed by the magnetic fields. 

We also carried out numerical simulations which included heat conduction 
starting from the same initial condition as \citet{hei03}.
We found that, when we adopt the Spitzer conductivity, the subcluster 
evaporates rapidly and the cold front is not formed.  
When we included magnetic fields and the effects of anisotropic heat 
conduction, the numerical results resembled Heinz's results. 
As suggested by \citet{hei03}, after the passage of shock waves, 
the cold plasma pulled back by gravity helps sharpen the cold front. 

In \S \ref{sec:m}, we presented numerical results that showed the 
K-H instability does not grow near the cold front, 
because the velocity of inflowing plasmas is reduced after the plasmas pass through 
the bow shock. In the magnetohydrodynamical models, the magnetic fields, which stretched 
along the contact surface, also stabilized the K-H instability. 
Thus, the K-H instability does not threaten the existence of 
cold fronts. 

In \S \ref{sec:mtc} we showed results of a model with initially disordered 
magnetic fields. Numerical results show that the magnetic fields near 
the front are gradually stretched. Thus, the initially tangled magnetic field 
lines become parallel to the front and suppress the heat conduction 
across the front. We confirmed the suggestion by \citet{vik01a} that 
the stretching of turbulent intergalactic magnetic fields enables 
the maintenance of cold fronts.
If the scale of magnetic turbulence is smaller than the mean free path, 
heat can be conducted almost isotropically. 
The heat conductivity in such a turbulent MHD medium 
is as large as $1/5$ of the Spitzer value \citep{nar01} and still large 
enough to diffuse out the cold front. We expect that coherent fluid motions 
and the stretching of the magnetic fields along the cold front increase 
the coherent length of field lines and make that region 
less turbulent. 

Let us now discuss the three-dimensional (3D) effects. Since magnetic fields 
compressed in front of the cold front can expand in the third direction, 
the strength of the magnetic fields in this region will be reduced in 3D. 
However, even such weakened magnetic fields suppress the heat conduction 
across them. Thus, we expect that cold fronts can be sustained even in 3D. 
We hope to report the results of such 3D MHD simulations in subsequent 
papers.

In this Letter, we assumed that the ambient plasma has magnetic fields. 
Recently, \citet{oka03} pointed out the possibility that 
magnetic fields are spontaneously generated by the Ramani-Laval instability 
around the cold front. Such magnetic fields can also help to suppress 
the heat conduction across the front.

\acknowledgments 

We thank T. Hanawa, S. Miyaji, N. Okabe, M. Hattori, and U. Fujita 
for their discussions.
The simulation code CANS that we used was developed as 
a product of the ACT-JST project (PI. R. Matsumoto). 
We thank T. Yokoyama for code development.
Numerical computations were carried out 
by VPP5000 at the Astronomical Data Analysis Center of 
the National Astronomical Observatory Japan (PI. N. Asai, project-ID yna23).

\clearpage

\begin{deluxetable}{cccc}
\tabletypesize{\scriptsize}
\tablecaption{Simulation models \label{tbl-1}}
\tablewidth{0pt}
\tablehead{
\colhead{Model} & \colhead{$\beta_{0}$} & \colhead{$\kappa$} 
& \colhead{simulation box}
}
\startdata
H & $\infty$  & $0$ & $2.5 \, {\rm Mpc} \times 2.5 \, {\rm Mpc}$\\
M & 100  & $0$ & $2.5 \, {\rm Mpc} \times 2.5 \, {\rm Mpc}$\\
HC & $\infty$ & $\kappa$ & $2.5 \, {\rm Mpc} \times 2.5 \, {\rm Mpc}$\\
MC &100 & $\kappa_{\parallel}$ & $2.5 \, {\rm Mpc} \times 2.5 \, {\rm Mpc}$\\
MTC & nonuniform & $\kappa_{\parallel}$& $7.5 \, {\rm Mpc} \times 2.5 \, {\rm Mpc}$\\
\enddata
\end{deluxetable}

\begin{figure}
\caption{
The initial condition of models M and MC. The color contour in the 
left panel shows the density. Arrows show the velocity vectors 
and white lines show the magnetic field lines. The right panel shows the 
distribution of density (solid curve), pressure (dotted curve), 
and temperature (dashed line) along $y = 0$.
The plasma beta in the ambient plasma is $\beta_{\rm 0} = 100$. 
\label{fig1}}
\end{figure}
\begin{figure}
\caption{
The effects of heat conduction and magnetic fields. 
The panels show the distributions of density, temperature, and  X-ray 
intensity at $t=3.0$ from left to right. The upper panels show 
the results for model MC (with magnetic fields and heat conduction), 
the lower panels show the results for model HC 
(without magnetic fields and  with heat conduction). 
The X-ray intensity distribution is visualized from the simulation 
results as the logarithm of $\rho^{2}$. 
The white curves and arrows show the magnetic field 
lines and velocity vectors, respectively. \label{fig2}} 
\end{figure}
\begin{figure}
\caption{
Results of the simulations without heat conduction 
for models M (left, with magnetic fields) and 
H (right, without magnetic fields). 
The color contours show the temperature at $t=3.0$. 
The solid curves and arrows show 
the magnetic field lines and velocity vectors, respectively. 
\label{fig3}} 
\end{figure}
\begin{figure}
\caption{
Time evolutions of model MTC (with disordered magnetic fields) 
at the central region of $5.0 \, \rm Mpc \times 2.5 \, \rm Mpc$ 
at $t =0$ and $t=3.0$. The top panels show the density distribution 
and the bottom panels show the temperature distribution. 
The white curves and arrows show the magnetic field lines 
and velocity vectors, respectively. \label{fig4}} 
\end{figure}

\end{document}